\documentclass[final,5p,times]{elsarticle}

\usepackage{amssymb}
\usepackage{amsmath}

\usepackage{hyperref}
\journal{Physics Letters A}

\usepackage[dvipsnames]{xcolor}
\usepackage{xspace}
\usepackage{bm}
\usepackage{siunitx}
\usepackage{upgreek}

\newcommand\md{\ensuremath{\mathrm{d}}\xspace}

\newcommand\GQS{\ensuremath{\mathrm{g}}\xspace}

\newcommand\Ai{\ensuremath{\mathrm{Ai}}\xspace}
\newcommand\Bi{\ensuremath{\mathrm{Bi}}\xspace}
\newcommand\cI{\ensuremath{\mathcal{I}}\xspace}

\newcommand\vc{\ensuremath{v_1^\ast}\xspace}
\newcommand\zc{\ensuremath{z_1^\ast}\xspace}
\newcommand\Zc{\ensuremath{Z^\ast}\xspace}
\newcommand\vcz{\ensuremath{v_0^\ast}\xspace}

\begin{document}
\begin{frontmatter}


\title{Image Theory for the Single Bounce Quantum Gravimeter}

\affiliation[LKB]{organization={Laboratoire Kastler Brossel,Sorbonne Universit{\'{e}}, CNRS, ENS-Universit{\'{e}} PSL, Coll{\`{e}}ge de France},
            addressline={4 place Jussieu},
            city={Paris},
            postcode={75005},
            state={},
            country={France}}

\author[LKB]{Joachim Guyomard} 
\author[LKB]{Serge Reynaud}
\author[LKB]{ Pierre Clad\'e}

\begin{abstract}
We develop an image theory for the recently proposed single-bounce quantum gravimeter. Free fall and quantum bounce of a matter wave-packet are described through decompositions over a basis of continuous energies. 
This leads to a much clearer interpretation of the origin of quantum interferences, associated to semi-classical estimations. 
We then give new tools to explore the space of parameters, and discuss the expected accuracy of the free-fall acceleration measurement.
\end{abstract}


\begin{keyword}



 
\end{keyword}

\end{frontmatter}

\section{Introduction}

The Universality of Free Fall \cite{Will2014llr} has been tested with high accuracy on macroscopic  \cite{Wagner2012,Viswanathan2018,Touboul2022prl,Touboul2022cqg} and atomic \cite{Herrmann2012cqg,Barrett2015njp,Asenbaum2020,Tino2020ppmp} test bodies. Ambitious projects are developed at CERN to test it on antihydrogen atoms \cite{Huber2000asr,Bertsche2015,bertsche2018,Yamazaki2020,Anderson2023} in order to gain information on the puzzle of the asymmetry observed between matter and antimatter in the Universe.
In particular, the aim of the GBAR experiment is to  push the accuracy of this test down to the $1\%$ level by timing the classical free fall of cold antihydrogen atoms \cite{Indelicato2014,Perez2015,Adrich2023}.

This accuracy can be improved by several orders of magnitude by using interferometry to measure quantum free fall of the atoms \cite{Crepin2019pra,Rousselle2022epjd}. The idea is to use quantum reflection of freely-falling atoms on a mirror \cite{Shimizu2001,Druzhinina2003,Oberst2005,Pasquini2006,Voronin2005jpb,Dufour2013,Crepin2017epl} to produce matter-wave interferences and measure the acceleration.
Even after a single bounce, the improvement is impressive with respect to the classical timing experiment  \cite{Guyomard2025}. 
This result is encouraging for the prospect of better accuracy for free-fall measurements on antihydrogen.
It also opens new perspectives for measuring gravity properties of rare or exotic species, of interest when intrinsic physical reasons limit the number of detection events or the time available for measurement.

The single-bounce design leads to a simple interference pattern, and then to a clearer comparison of theory and experiment. It  is quite different from atomic gravimeters commonly used to measure free fall, where interferences are produced between waves having followed different trajectories after beam splitters \cite{kasevich1991,Storey1994,Borde2001,Gillot2014}. It is therefore interesting to gain a better understanding of the principle of this new tool, as well as of the optimisation of parameters which determine its performances.

It is the aim of this paper to present a new analysis of the single-bounce quantum gravimeter.
Matter waves are now expanded over a continuous basis of freely-falling eigen-solutions while the bounce on a perfect mirror is described as an energy-dependent scattering (\S \ref{sec:freefall}). The new analysis combines effects of free fall and bounce and the far-field approximation is used to understand its results  (\S \ref{sec:farfield}). A semi-classical estimation is given for the phase of the interference pattern, in both cases of far-field and not-so-far-field propagation (\S \ref{sec:semiclassical}). A simple formula is given for the expected accuracy which is an approximation of the results obtained from exact results (\S \ref{sec:discussions}).

\section{Free fall and quantum bounce}
\label{sec:freefall}

Freely falling matter waves may be developed on a basis of solutions of the Schrödinger equation. This is usually done on the discrete basis of gravitational quantum states (GQS), the eigen-states of the cavity formed by gravity and reflection on the mirror \cite{Breit1928,Wallis1992,Nesvizhevsky2002nature,Nesvizhevsky2003,Nesvizhevsky2005,Jenke2011,Pignol2014,Nesvizhevsky2015}. Here we use a continuous basis of solutions corresponding to real energies, which will allow us to develop a more intuitive understanding of the phase of the interference pattern. 

Any freely falling wave is a superposition of elementary solutions of Schrödinger equation
\begin{equation}
\begin{aligned}
&\Psi(z,t)=\int_{-\infty}^\infty c(E)\,{\psi_E(z)}\,
e^{\frac{-\imath Et}{\hbar}} \, {\md E}
~,\\
&\quad c(E)= \int_{-\infty}^\infty \Psi(z,0)\,\psi_E(z)\,  {\md z}~.\\
\end{aligned}
\label{eq:freefall}
\end{equation}
Waves are described by wave-functions $\Psi$ or amplitude $c$ equivalently, with appropriate normalizations
\begin{equation}
\begin{aligned}
&\quad\int{\psi}_E(z){\psi}_{E^\prime}(z) 
\,\md z= \delta\left(E-E^\prime\right)  ~,\\
&\int \left|c(E)\right|^2 \md E = 1 
\;\Rightarrow \;
\int \left|\Psi(z,t)\right|^z \md z = 1   ~.
\end{aligned}
\label{eq:normalizations}
\end{equation}
Elementary solutions are Airy function of arguments reduced to natural scales, a length $\ell_\GQS$ and an energy $e_\GQS$.
\begin{equation}
\begin{aligned}
\psi_E(z)&=\frac{\Ai\left(\zeta-\epsilon\right)}{\sqrt{\ell_\GQS\, e_\GQS}}
\;,\quad  \zeta=\frac{z}{\ell_\GQS}
\;,\quad  \epsilon=\frac{E}{e_\GQS}  ~,\\
\quad\ell_\GQS&=\sqrt[3]{\frac{\hbar^2}{2gm^2}}
\;,\quad  e_\GQS=mg\ell_\GQS  ~.
\end{aligned}
\label{eq:elementary}
\end{equation}
This implies that representations $\Psi$ and $c$ are related to each other through Airy transforms \cite{Vallee2004}.

The two representations are available before and after the bounce on the mirror, for waves labelled respectively by subscripts $i=0,1$ in $\Psi_i$ or $c_i$. 
The bounce is described by a scattering amplitude $\rho(E)$ and simply written in energy representation 
\begin{equation}
c_1(E) = \rho(E)\,{c_{0}(E)} ~.
\label{eq:scattering}
\end{equation}

We consider for simplicity a perfectly reflecting mirror, so that the amplitude $\rho(E)$ is a complex number with unit modulus. Its phase is calculated from continuity conditions on the quantum reflection mirror.
As the sum of incoming downward wave and outgoing upward wave vanishes on the mirror at $\zeta=0$, one gets
\begin{equation}
\rho(E)=
-\frac{\Ai\left(-\epsilon\right)-\imath\, \Bi\left(-\epsilon\right)}{\Ai\left(-\epsilon\right)+\imath\, \Bi\left(-\epsilon\right)} =e^{2\imath\pi k\left({E}\right)}~.
\label{eq:reflectionamplitude}
\end{equation}
The scattering phase $2\pi k\left({E}\right)$ is related to the phase of Airy functions defined as in \S 9.8 in \cite{DLMF}.
It is increasing as $E$ increases, and behaves asymptotically as $\frac43 \left(\frac{E}{e_\GQS}\right)^{3/2}$ at large positive $E$ (eq.~9.8.22 in \cite{DLMF}). 

\begin{figure}
    \centering
    \includegraphics[viewport = 35 15 260 130, clip, width=1\linewidth]{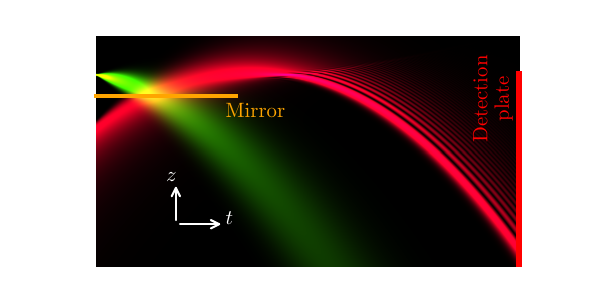}
    \caption{Squared modulus of the initial wave-function $\Psi_0$ (in green) and of the reflected wave-function $\Psi_1$ (in red) shown as density plots versus altitude $z$ and time $t$.  The orange line represents the mirror plate at altitude $z=0$.
    }
    \label{fig:sourceandimage}
\end{figure}


For explicit drawings shown below, we specify $\Psi_0$  as a Gaussian function of $z$ with noticeable values only above the mirror. 
Relevant parameters are the mean altitude $z_0$ of the wave-packet above the mirror plate, the mean vertical velocity $v_0$ and the altitude dispersion $\sigma_z$, assumed to be much smaller than $z_0$. 
This choice, the same as in \cite{Guyomard2025}, leads to an analytical expression for $c_0$ 
\begin{equation}
\begin{aligned}
&c_0(E) = \sqrt[4]{8\pi} \sqrt{\frac{\sigma_\zeta }{e_\GQS}} \;
   \Ai\left(\zeta_0 - \epsilon+\sigma_\zeta^2 (\imath \nu_0 + \sigma_\zeta^2) \right)    \\
& \quad\times\exp{\left(\sigma_\zeta^2 \left(\zeta_0 - \epsilon
- \frac{\nu_0^2}4 + \sigma_\zeta^2 \left(\imath \nu_0  + \frac{2\sigma_\zeta^2}3 \right)\right)\right)}
 ~,    \\
&\zeta_0=\frac{z_0}{\ell_\GQS} \;,\quad  
\nu_0=\frac{v_0\,t_\GQS}{\ell_\GQS}\; ,\quad 
t_\GQS=\frac{\hbar}{e_\GQS} \; ,\quad 
\sigma_\zeta=\frac{\sigma_z}{\ell_\GQS} ~.
\end{aligned}    
\end{equation}
The discrete amplitudes $c_n$ in \cite{Guyomard2025} are given by $c(E)$ at the discrete energies $E_n$ corresponding to GQS. 

The core idea of the image theory is that all information about the reflected wave is described exactly by the image wave-function $\Psi_1$ which experiences free-fall evolution after the bounce 
\begin{equation}
\Psi_1(z,t)=\int_{-\infty}^\infty c_1(E)\, \psi_E(z)\;
e^{\frac{-\imath Et}{\hbar}} \, {\md E }  ~.
\label{eq:imagewave}
\end{equation}
The square moduli of source and image wave-functions $\Psi_0$ and $\Psi_1$  are drawn as density plots on Fig.\ref{fig:sourceandimage}, showing the effect of the bounce. 

We can consider $\Psi_1$ as the result of free-fall evolution from the (virtual) wave-function $\Psi_1$ evaluated at $t=0$.
The bounce can be understood as transforming the source wave $\Psi_0$ into the image wave $\Psi_1$ at $t=0$. 
The propagation of the image wave is then described by the free-fall propagator, described for example in \cite{Storey1994} or \cite{Guyomard2025}.
As the time of entrance in the gallery and the conserved horizontal velocity $V$ are supposed for simplicity to have no dispersion, the horizontal position and the time play equivalent roles.

\begin{figure*}
    \centering
    \includegraphics[width=1\linewidth]{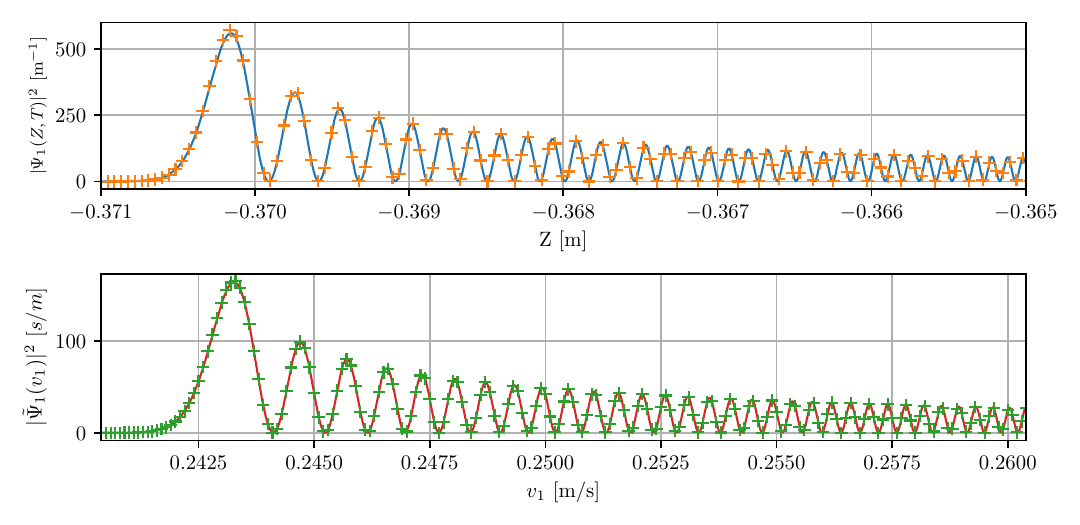}
    \caption{ Top plot : Square modulus $\vert{\Psi}_1(Z,T)\vert^2$ of the exact wave-function (blue solid line) and its model to be described in \autoref{sec:semiclassical} {(orange crosses)}, drawn versus position $Z$ at detector, with the parameters used for Fig.3 in \cite{Guyomard2025}: $z_0=\SI{1}{\milli\meter}$, $v_0 = \SI{-91.5}{\milli\meter\per\second}$, $\sigma_v = \SI{79}{\milli\meter\per\second}$, $T=\SI{300}{\milli\second}$. Bottom plot: Square modulus $\vert\widetilde{\Psi}_1(v_1,0)\vert^2$ of the image wave-function (red solid line) and its model to be described in \autoref{sec:semiclassical} {(green crosses)}, drawn versus velocity $v_1$ at $t=0$.}
\label{fig:comparisonsquaremoduli}
\end{figure*}

The signal is the probability distribution $\vert\Psi_1(Z,T)\vert^2$ at vertical position $Z$ on the detector, at the detection time $T=\frac{D}{V}$ ($D$  the horizontal distance from the source to the detector).
This distribution is drawn as the blue line on the top plot in Fig.\ref{fig:comparisonsquaremoduli}, for the values of parameters used for Fig.3 in  \cite{Guyomard2025} ({$z_0=1\,\mathrm{mm}$, $v_0=-91.5\,\mathrm{mm/s}$, $\sigma_z\simeq0.4\,\mathrm{\mu m}$} and $T=300\,\mathrm{ms}$). 
It is identical to the distribution obtained from calculations based on the discrete decomposition on GQS in \cite{Guyomard2025}.

The Fisher information $\cI_Z$ calculated from this distribution determines the Cramer-Rao bound $\sigma_\mathrm{CR}$ for the accuracy expected in the limit of a high enough number $N$ of detected events, when precision is limited by statistics \cite{Frechet1941,Cramer1946,Refregier2004}
\begin{equation}
\begin{aligned}
\cI_Z& =  \left(2g_0\right)^2 \int \left(\partial_g \vert\Psi_1(Z,T)\vert\right)^2 \,\md Z  ~,\\
&\frac{\sigma_\mathrm{CR}}{g_0} = \frac{1}{\sqrt{N\,\cI_Z}} ~.
\end{aligned}
\label{eq:fisherZ}
\end{equation}
The reference value $g_0$, taken for example as the standard free-fall acceleration ar Earth surface, has been inserted in order to get dimensionless numbers in \eqref{eq:fisherZ}.

\section{Far-field approximation}
\label{sec:farfield}

The arguments in the previous section suggest a new route to a deeper understanding of the results, that we follow now.
The idea is to look at the limit of a long free fall time $T$, where the detection position $Z$ is determined by the momentum $p_1$ in the image wavefunction at initial time $t=0$. 
This far-field approximation leads to a simple interpretation of the origin of the interference pattern seen on the detector. 

In the far-field approximation, one gets a simplified form for the signal with a one-to-one mapping of the final position $Z$ to the initial momentum $p_1$ or velocity $v_1=\frac{p_1}{m}$  and a factor corresponding to the Jacobian in the change of variables
(using momentum is natural as Fourier transforms are involved, while using velocity is more convenient for drawings and mechanical intuition)
\begin{equation}
\begin{aligned}
&\vert{\Psi}_1(Z,T)\vert^2  \simeq \frac mT \,
\vert\widetilde{\Psi}_1(p_Z,0)\vert^2  ~, \\
&\; p_Z= \frac mT\left(Z+\frac{g\,T^2}{2}-\zc \right) ~.
\end{aligned}
\label{eq:positionprobapprox}
\end{equation}
Here, $\zc$ is a reference position in the image wave at $t=0$, the expression of which will be given in eq.\ref{eq:brancpoint}. Meanwhile, $\widetilde{\Psi}_1(p,0)$ is the image wavefunction in momentum space at $t=0$, that is  the Fourier transform of the image wavefunction in position space (using velocity instead of momentum only changes the normalization of ; $\vert\widetilde{\Psi}_1\vert$by a constant factor $\sqrt{m}$)
\begin{equation}
\begin{aligned}
&\; \widetilde{\Psi}_1(p,0)= \int_{-\infty}^\infty 
c_1(E)\, {\widetilde{\psi}_E(p)} \,{\md E } ~, \\
&\widetilde{\psi}_E(p) = 
\frac{1}{\sqrt{2\pi\hbar mg}} \, \exp\left( - \frac{\imath  p}{\hbar mg}\left(E - \frac{p^2}{6m}\right)\right) ~.
\end{aligned}
\label{eq:Psitilde}
\end{equation}

The distribution in this far-field approximation is drawn as the orange line on the bottom plot in Fig.\ref{fig:comparisonsquaremoduli}), with the same parameters.
The two plots look very much alike which means that the main features of the interference pattern are captured in the approximation. Precise comparison shows slight differences in the positions of the fringes, due to the fact that the value of $T$ is not large enough for the far-field approximation to be valid. This qualitative discussion will be made more precise below after a semi-classical estimation of the phases. 

We now understand the improved measurement of the free-fall acceleration as due to the fine structures in momentum $p_1$ imprinted on the image wavefunction by the phase of the reflection amplitude $\rho(E)$. These structures in $p_1$ are revealed after propagation as structures in $Z$ on the detection screen.
Let us note that the classical experiment without mirror would correspond to an initial wavefunction $\widetilde{\Psi}_0$ being a Gaussian function in $p_0$. This clearly explains how the measurement is improved when going from the classical version to the quantum one.
 
As the new approach produces better insight on the origin of the interferences, we will use it in the following to estimate the expected accuracy.
The Fisher information $\cI_Z$ will thus be approximated by another integral now calculated from the $p-$distribution
\begin{equation}
\begin{aligned}
\cI_p=\int \left(\left(2g_0\,\partial_g + mg_0T\,\partial_p \right)
\vert\widetilde{\Psi}_1(p,0)\vert\right)^2 \,\md p  \,.
\end{aligned}
\label{eq:fisherp}
\end{equation}
The integrand is the square of a sum of two terms, with the first and second  terms representing the direct dependence of $\widetilde{\Psi}_1$ on $g$ and the effect of $p-$structures respectively. The second term is larger than the first one by a factor proportional to $T$. 

Variations with parameters of the two expressions are expected to be close to each other ($\cI_Z \simeq \cI_p$) when the far-field approximation is reliable. 
We will present more precise comparisons below by using models of the interference pattern. 
We stress at this point that calculations without any  approximation remain mandatory for exact evaluations and for data analysis of the results of any forthcoming experiment. 

\section{Feynman path analysis}
\label{sec:semiclassical}

Before entering the discussion of expected accuracy, we give  semi-classical estimations of the interference patterns. These estimations, based on a Feynman path analysis of the problem \cite{FeynmanHibbs1965}, will be useful as models for the exact and far-field expressions, and they will also help us to obtain a prediction for the expected accuracy in terms of the various parameters. 

\begin{figure}
    \centering
    \includegraphics[viewport = 35 15 260 130, clip, width=1\linewidth]{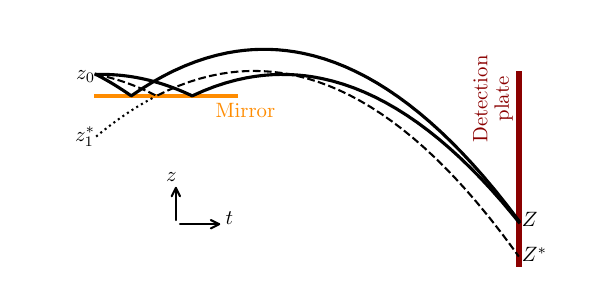}
    \caption{Classical trajectories associated with the interferometer described in \autoref{fig:sourceandimage}. The two solid lines represent classical trajectories going from the source $z_0,t=0$ to a detection point $Z,T$. The dashed line corresponds to the limit of coincidence of these two trajectories and arrival at the branchpoint $\Zc$. No classical trajectory reaches the detection plate at $Z$ below the branchpoint $\Zc$.}
    \label{fig:trajectoires_classiques}
\end{figure}

In the Feynman path analysis, we look at classical trajectories going from the source point at altitude $z_0$ and time $t=0$ to the detection point at altitude $Z$ and time $t=T$ after a single bounce on the mirror at altitude $z=0$. 
This analysis is illustrated by the semi-classical scheme in Fig.     \ref{fig:trajectoires_classiques}.
It produces a third degree equation for the time $t_b$ at which the bounce happens
\begin{equation}
z_0\left(T - t_b\right) +Z t_b+ g t_b\left(T - t_b\right)\left(\frac T2 - t_b\right) =0  ~ ,
\label{eq:3rddegree}
\end{equation}
that is also under a reduced form
\begin{equation}
\begin{aligned}
&\qquad \overline{t}_b^3 - \frac{3\lambda}{2g} \overline{t}_b - \frac{T\mu}{2g}  =0 \;,\quad \overline{t}_b = t_b-\frac T2 \,,\\
&L=\frac{gT^2}{2} \;,\quad
\lambda=\frac{L-2\left(Z+z_0\right)}{3}\;,\quad
\mu=z_0-Z \;.
\end{aligned}
\label{eq:reduced}
\end{equation}
For the parameters chosen for the interference pattern in Fig. \ref{fig:comparisonsquaremoduli}, this equation has two solutions corresponding to bounces on the mirror with ${t}_b$ positive and small compared to $T$ (the third solution is close to $3T/2$ and therefore not physically relevant). The interference peaks are close to the branchpoint where the two physical solutions coincide at a transition from the case where they are real to that where they are conjugated complex numbers. 

At the branchpoint the discriminant $D$  vanishes
 \begin{equation}
D=\lambda^3-L \mu^2 \;.
\label{eq:discri}
\end{equation}
The relevant branchpoint for our problem is the solution of the equation $D=0$ close to 
\begin{equation}
\begin{aligned}
&\quad \Zc  \simeq -L+\vc T + \zc \;,\\
&\vc =\sqrt{6gz_0} \;,\quad
\zc =-\frac{5z_0}3{} \;.
\end{aligned}
\label{eq:brancpoint}
\end{equation}
The velocity $\vc$ and position $\zc$ are initial values of $v_1$ and $z_1$ on the classical trajectory corresponding to arrival at the branchpoint, evaluated at far-field limit. 

In the vicinity of the branchpoint (for $D>0$), the difference $\Delta t_b$ of the physical solutions is proportional to $\sqrt{D}$. Meanwhile, the classical action is extremal on each classical path and its derivative with respect to time is proportional to $D$ near the branchpoint. Hence the difference $\Delta\Phi$ of phases calculated along the two paths arriving at position $Z$ varies as $D^{3/2}$ there. This implies that the phase of the interference pattern is the phase of the Airy function $\Ai$ of an argument proportional to $D$. 

Parameters have been chosen so that the fringe contrast is good, which means that the initial velocities $v_0$ for the two classical paths correspond to nearly equal Gaussian amplitudes in the initial velocity distribution. The reduction of the contrast due to the Gaussian distribution is then described by the exponential of an argument proportional to $D$ (difference $\Delta v_0$ of initial velocities scales as $\sqrt{D}$, as difference $\Delta t_b$ of bounce times).

These qualitative ideas can be made more precise, by first calculating the phase $\Delta\Phi$ in a semi-classical analysis and then using an Airy uniform approximation across the branchpoint \cite{Berry1972,Knoll1977} (see also \cite{Vallee2004}), which leads to 
\begin{equation}
\begin{aligned}
&\vert\Psi_1\left(Z,T\right)\vert \simeq 
\left(\frac{8\pi}{\gamma}\right)^{1/4} \,
\sqrt{\frac{\md \Delta}{\md Z}}
\, \vert\Ai\left(- \Delta\right) \vert
\, e^{-\frac{\Delta}{\gamma}} \,,\\
&\Delta\left(Z\right)=
\frac{m^{2/3}\,D}{\left(2\hbar T\sqrt{3L}\lambda \mu\right)^{2/3}}   
 \;,\quad
\gamma=\left(\frac{ 3m}{\hbar g} \right)^{2/3} \sigma_v^2 \;.
\end{aligned}
\label{eq:Delta}
\end{equation}
The expression of $\Delta$ has been fixed from the semi-classical phase $\Delta\Phi$ so that the model \eqref{eq:Delta} matches the interference pattern. Precisely the fringes in top plot of Fig.\ref{fig:comparisonsquaremoduli} are well reproduced over the whole range shown there, with the points given by the model lying on the exact curve calculated above for the same parameters. The expressions of $\gamma$ and of the factor in front of the distribution have been chosen to match the slow decay of the envelop. The dimensionless parameter $\gamma$ has to be understood as the number of fringes in the pattern, and the distribution \eqref{eq:Delta} is normalized under the condition $\gamma\gg1$ (which is assumed in the following). 

The model can be used as well for larger values of $T$ where the far-field approximation gives a good description. This implies that the exact expression obtained above for the initial image wavefunction can be given a good analytical model by taking the far-field limit of \eqref{eq:Delta}
\begin{equation}
\begin{aligned}
\vert\widetilde{\Psi}_1\left(v_1,0\right)\vert &\simeq
 \left(\frac{8\pi}{\gamma}\right)^{1/4}  \sqrt{\frac{\md\widetilde{\Delta}}{\md v_1}} 
 \,\vert\Ai\left(-\widetilde{\Delta}\right) \vert
\,e^ {-\frac{\widetilde{\Delta}}{{\gamma}}}  \,,\\
\widetilde{\Delta}\left(v_1\right)&=\frac{m^{2/3}\left(v_1^2-\vc\,^2\right)}{\left(9\hbar g\right)^{2/3}}   \,.
\end{aligned}
\label{eq:Deltafarfield}
\end{equation}
The agreement of the model \eqref{eq:Deltafarfield} with the exact result is very good, as shown on bottom plot in Fig. \ref{fig:comparisonsquaremoduli}, where the points (model) lie on the solid curve (exact pattern).  Note that the analytical model \eqref{eq:Deltafarfield}
 could also have been proven directly from the expression \eqref{eq:Psitilde}.

\section{Expected accuracy}
\label{sec:discussions}

We now use the results of the preceding section to produce a simple prediction of how the Fisher information and then the expected accuracy vary with parameters.

To this aim, we use the far-field expression \eqref{eq:fisherp} of the Fisher information as well as the analytical model \eqref{eq:Deltafarfield} of $\vert\widetilde{\Psi}_1\left(v_1,0\right)\vert$. We then note that the Fisher integral is obtained under closed form when the slowly varying derivatives of ${\widetilde{\Delta}}$ are approximated as constant.
This leads to a simple prediction for the Fisher information that will be used in the following 
\begin{equation}
\cI_S \simeq \frac{2 \left(mgz_0\right) \left(m \sigma_{v}^2\right) T^2}{3 \hbar^{2}} \,.
\label{eq:fisher_simple_feynmann}
\end{equation}
The expression has been written to be clearly dimensionless, as the product of two numbers each obtained as an energy multiplied by $T/\hbar$ (it could also be written as the product of length scales $z_0$ and $gT^2$ divided by $\sigma_z^2$).

In the sequel of this section, we compare this simple prediction $\cI_S$ for the Fisher information (eq. \ref{eq:fisher_simple_feynmann}) to the quantity ${\cI_Z}$ calculated on the $Z-$distribution at $t=T$ (eq.\eqref{eq:fisherZ}).
We note that the calculation of $\cI_Z$ is simpler to implement numerically with this model than with the discrete GQS basis as done in \cite{Guyomard2025}. Indeed, the distribution in $Z$ of the wave function at $T$ is, up to a phase, the Fourier transform of the distribution in $p_1$ of the wave function at $t=0$ and the latter is also, up to a phase, the Fourier transform of the amplitude $c$ (see \autoref{eq:Psitilde}), both of which can be calculated using FFT.


\begin{figure}[th]
    \centering
    \includegraphics[width=1\linewidth]{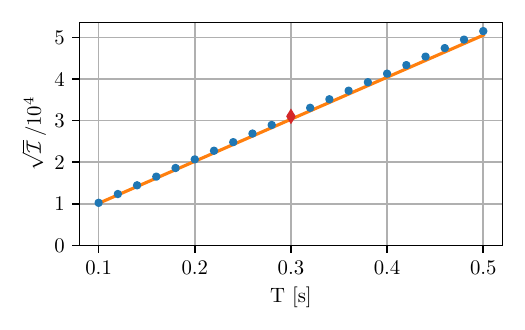}
    \caption{Variation of square root of Fisher information versus propagation time $T$, with all other parameters the same as for Fig.\ref{fig:comparisonsquaremoduli}. The simple prediction $\sqrt{\cI_S}$ (solid orange line) and the exact quantity $\sqrt{\cI_Z}$  (blue dots) agree reasonnably well, and the accuracy is improved when increasing $T$ (the red diamond corresponds to the value $T=300\,\mathrm{ms}$ chosen for  Fig.\ref{fig:comparisonsquaremoduli}).  }
    \label{fig:variationT}
\end{figure}

Figure  \ref{fig:variationT} represents the variations of square roots $\sqrt{\cI_S}$ (orange solid curve) and $\sqrt{\cI_Z}$ (blue dots)  when the propagation time $T$ is varied, with all other parameters kept fixed to the values chosen for Fig.\ref{fig:comparisonsquaremoduli}.  
The figure first shows the good agreement between the two evaluations, in spite of the crude assumptions made in the calculation of $\cI_S$. It also shows that $\cI$ is increased, and therefore the expected accuracy improved, when the total time of flight $T$ is larger and larger. This had of course to be expected for a free-flight measurement, but the prediction \eqref{eq:fisher_simple_feynmann} now provides us with an estimation of the improvement.

\begin{figure}[th]
    \centering
    \includegraphics[width=1\linewidth]{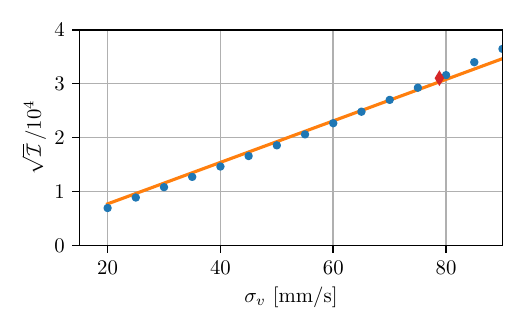}
    \caption{Variation of square root of Fisher information versus initial velocity dispersion $\sigma_v$, with all other parameters the same as for Fig.\ref{fig:comparisonsquaremoduli}. The simple prediction $\sqrt{\cI_S}$ (solid orange line) and the exact quantity $\sqrt{\cI_Z}$  (blue dots) agree reasonnably well, and the accuracy is improved when increasing $\sigma_v$ (the red diamond corresponds to the value $\sigma_v = \SI{79}{\milli\meter\per\second}$ chosen for  Fig.~\ref{fig:comparisonsquaremoduli}). The upper value for $\sigma_v$ is chosen so that the whole initial distribution corresponds essentially to one and only one bounce on the mirror, as discussed in \cite{Guyomard2025}.}
    \label{fig:variationsigmav}
\end{figure}

We then illustrate in Fig. \ref{fig:variationsigmav} a similar discussion of the variations obtained when varying the initial velocity dispersion $\sigma_v$.  The figure again shows the good agreement of the two evaluations. It also shows the expected accuracy is improved when the initial velocity dispersion $\sigma_v$ is increased. The interpretation of this observation is that $\gamma$ scales with $\sigma_v^2$ so that the number of fringes contributing to the measurement is increasing as $\sigma_v^2$. 
Note that the simple prediction would go to infinity in the absence of the exponential decay superimposed to the decay of Airy function in the models \eqref{eq:Delta} and \eqref{eq:Deltafarfield}.
This might be seen as counter-intuitive at first sight. It is however understood from the fact that each added fringe brings a similar information due to its more and more rapid variation which compensates its lower and lower amplitude. This raises the question of the better and better spatial resolution required on the detector to take benefit of finer and finer fringes (see the detailed discussion dedicated to this point in \cite{Guyomard2025}). 

\begin{figure}[th]
    \centering
    \includegraphics[width=1\linewidth]{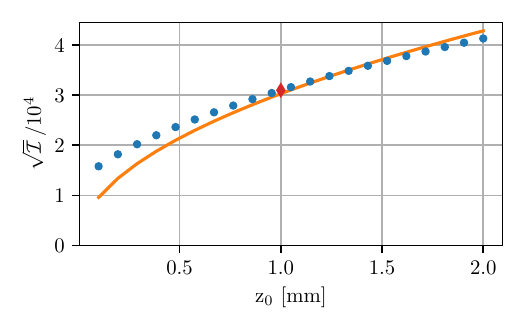}
    \caption{Variation of square root of Fisher information versus the initial height $z_0$ above the mirror, with $v_0 = -\frac{\sqrt{6gz_0}}{3}$ and other parameters the same as for Fig.~\ref{fig:comparisonsquaremoduli}. The simple prediction $\sqrt{\cI_S}$ (solid orange line) and the exact quantity $\sqrt{\cI_Z}$  (blue dots) are roughly agreeing, indicating that the accuracy is improved when $\sigma_v$ is increased (the red diamond corresponds to the value $z_0=1\,\mathrm{mm}$ chosen for  Fig..\ref{fig:comparisonsquaremoduli}).}
    \label{fig:variationz0}
\end{figure}

We end up the discussions in the present section by showing in Fig. \ref{fig:variationz0} the variations of $\cI$ when the initial height $z_0$ above the mirror is varied. 
This requires to discuss also the fact that the initial mean velocity $v_0$ has to be varied when $z_0$ is varied, because we want the fringes to keep a good contrast.  
The condition is easily written at the far-field limit 
\begin{equation}
v_0 = \vcz =-\frac{\sqrt{6gz_0}}{3} \,,
\label{eq:optimumv0}
\end{equation}
where $\vcz$ is the initial values of $v_0$ of the classical trajectory corresponding to arrival at the branchpoint, evaluated at far-field limit.
The condition \eqref{eq:optimumv0} amounts to keep a constant ratio between mean kinetic and potential energies in the initial quantum state. 

Note that the choice \eqref{eq:optimumv0} is only an approximation when the far-field limit is not exact, which is the case for the parameters chosen for drawing Fig.\ref{fig:comparisonsquaremoduli}. This is why we decided to optimize  $\cI_Z$ in \cite{Guyomard2025} by chosing the specific value $v_0=-91.5\,\mathrm{mm/s}$, whereas \eqref{eq:optimumv0} would have led to a different value {$v_0=-87\,\mathrm{mm/s}$} (it would be the optimum for the exact expression $\cI_Z$ at the far-field limit $T\to\infty$). This precision is important for comparing results in \cite{Guyomard2025} and in the present paper, but it does not modify the broader discussion in the present section.

For Fig. \ref{fig:variationz0}, we have drawn variations of $\cI$, when $z_0$ is varied and $v_0$ given by \eqref{eq:optimumv0}, with other parameters kept fixed to  the values chosen for Fig.\ref{fig:comparisonsquaremoduli}. 
The agreement between the two methods is less good here but it clearly indicates a trend to an improvement of the accuracy when $z_0$ is increased, again consistent with expecation for a free fall measurement.
The upper value for abscissa $z_0$ on Fig. \ref{fig:variationz0} is chosen so that the losses due to imperfect quantum reflection remain limited (see the detailed discussion dedicated to this point  in \cite{Guyomard2025}). 

\section{Conclusions}

We have presented a new analysis of the recently proposed concept of single bounce quantum gravimeter \cite{Guyomard2025}. This has led us to a better understanding of the origin of interferences already present in momentum representation of the image wave-function, that is  the reflected wave.
We also obtained an analytical model of the position distribution at the detection plate, and deduced a simple estimation of the 
expected accuracy describing its variation when the parameter space is explored. 
This simple estimation will be useful to evaluate the feasibility of applications of the idea for investigating gravitational properties of rare or exotic species, for which number of events or time available for measurement would be limited for basic  reasons.
 
The reflection amplitude \eqref{eq:reflectionamplitude} has been given for perfect quantum reflection with the aim of better understanding the results of  \cite{Guyomard2025}. We may however stress here that an advantage of the new method (not yet exploited in this paper) is to make it easier to use the already available knowledge of quantum reflection on the Casimir-Polder potential  \cite{Crepin2017pra,Crepin2019hi,Crepin2019epjd}.
In this way, we would be able to account for the losses on the bouncer, with appropriate changes in the equations. 

\section*{Acknowledgments}
We thank our colleagues in  GBAR https://gbar.web.cern.ch/ and 
  GRASIAN  https://grasian.eu/ collaborations for
 insightful discussions, in particular S. Baessler, C. Blondel, 
 P. P. Blumer, C.  Christen, P.-P. Crepin, P. Crivelli, P. Debu, 
 A. Douillet, C. Drag, N. Garroum, R. Guérout, L. Hilico, 
 P. Indelicato, G. Janka, J.-P. Karr, S. Guellati-Khelifa, L. Liszkay, 
 B. Mansoulié, V. V. Nesvizhevsky, F. Nez, N. Paul, P. Pérez, 
 J. Pioquinto, C. Regenfus, O. Rousselle, F. Schmidt-Kaler, 
 K. Schreiner, A. Yu. Voronin, S. Wolf, P. Yzombard. 
 This work was supported by the Programme National GRAM of CNRS/INSU with INP and IN2P3 co-funded by CNES, and by 
 Agence Nationale pour la Recherche, Photoplus project 
 Nr. ANR 21-CE30-0047-01. The PhD work of Joachim Guyomard was supported by QuantEdu-France (ANR-22-CMAS-0001) in the framework of France 2030.

\bibliographystyle{elsarticle-num} 
\bibliography{bibliography}



\end{document}